\documentclass[a4paper,fleqn,usenatbib,useAMS]{mnras}

\usepackage{graphicx}	
\usepackage{amsmath}	
\usepackage{amssymb}	
\usepackage{multicol}        
\usepackage{bm}		
\usepackage{pdflscape}	

\usepackage[T1]{fontenc}
\usepackage{ae,aecompl}
\usepackage{mathptmx}

\usepackage{hyperref}
\PassOptionsToPackage{hyphens}{url}\usepackage{hyperref}
\makeatletter
\g@addto@macro{\UrlBreaks}{\UrlOrds}
\makeatother

\newcommand{\kms}{\,km\,s$^{-1}$} 

\title[Where have all the low-metallicity galaxies gone?]{Where have all the low-metallicity galaxies gone? Tracing evolution in the mass--metallicity plane since a redshift of 0.7}

\author[S. Zhou]{Shuang Zhou$^{1}$\thanks{Contact e-mail: \href{mailto:Shuang.Zhou@nottingham.ac.uk}{Shuang.Zhou@nottingham.ac.uk}},
Alfonso Arag{\'o}n-Salamanca$^{1}$,
Michael Merrifield$^{1}$, V. M. Sampaio$^{1,2}$
\\
$^{1}$School of Physics \& Astronomy, University of Nottingham, University Park, Nottingham, NG7 2RD, UK\\
$^{2}$ NAT - Universidade Cruzeiro do Sul /  Universidade Cidade de S\~ao Paulo, 01506-000, SP, Brazil\\
}
\date{Last updated ???; in original form ???}

\pubyear{2022}

\begin{document}

\label{firstpage}
\pagerange{\pageref{firstpage}--\pageref{lastpage}}
\maketitle

\begin{abstract}
Even over relatively recent epochs, galaxies have evolved significantly in their location in the mass--metallicity plane, which must be telling us something about the latter stages of galaxy evolution. In this paper, we analyse data from the LEGA-C survey using semi-analytic spectral and photometric fitting to determine these galaxies' evolution up to their observed epoch at $z \sim 0.7$. We confirm that, at $z \sim 0.7$, many objects already lie on the present-day mass--metallicity relation, but with a significant tail of high-mass low-metallicity galaxies that is not seen in the nearby Universe. Similar modelling of the evolution of galaxies in the nearby MaNGA survey allows us to reconstruct their properties at $z \sim 0.7$.  Once selection criteria similar to those of LEGA-C are applied, we reassuringly find that the MaNGA galaxies populate the mass--metallicity plane in the same way at $z \sim 0.7$.  Matching the LEGA-C sample to their mass--metallicity ``twins'' in MaNGA at this redshift, we can explore the likely subsequent evolution of individual LEGA-C galaxies. Galaxies already on the present-day mass--metallicity relation form few more stars and their disks fade, so they become smaller and more bulge-like.  By contrast, the high-mass low-metallicity galaxies grow their disks through late star formation, and evolve rapidly to higher metallicities due to a cut-off in their wind-driven mass loss.  There are significant indications that this late cut-off is associated with the belated end of strong AGN activity in these objects.

\end{abstract}

\begin{keywords}
galaxies: fundamental parameters -- galaxies: stellar content --galaxies: formation -- galaxies: evolution
\end{keywords}


\section{Introduction}
The evolution of galaxies will alter both their masses, as more stars form, and metallicities, as successive generations of stars pollute their hosts with heavy elements.  We might therefore expect the position of a galaxy in the mass--metallicity plane to tell us something about its formation, and changes in the way this plane is populated over time can be used as an indicator of evolution in the galaxy population.

Since star formation couples mass and metallicity evolution, one might expect these two quantities to be correlated.  Indeed, as long ago as \cite{Lequeux1979} it was known that galaxies lie on a relatively tight mass--metallicity relation (hereafter MZR, since metallicity is usually quantified with the symbol $Z$).  The Sloan Digital Sky Survey (SDSS) vastly expanded the data set available, and allowed the MZR to be explored using measures of metallicity derived from both the gas phase \citep[e.g.][]{Tremonti2004} and the integrated starlight \citep[e.g.][]{Gallazzi2005}, both of which show a fairly tight relation in nearby galaxies across several orders of magnitude in mass.  

To seek clues as to how the MZR was established, one can compare results in the nearby Universe to those found at significant look-back times. Indeed, one does not have to look back very far -- even at relatively modest redshifts well below one, not all galaxies seem to follow the present-day MZR \citep[e.g.][]{Zahid2011,Maier2015,Huang2019, Lewis2023}. Studying the gas-phase metallicities of galaxies in the zCOSMOS survey, \cite{Maier2015} found that, while many intermediate-redshift high-mass galaxies at $z\sim0.7$ already lie on the present-day MZR, there was a sub-population of ``massive low-metallicity galaxies'' that were metal deficient by a factor of 2--3. Similar results are seen in the recent measurements by \cite{Lewis2023}, such that some of the massive galaxies at $z\sim0.7$  can have metallicity up to 0.4 dex lower than the present-day MZR. Unfortunately, the interpretation of relatively limited emission-line data in terms of metallicities is complicated by the effects of dust obscuration and contamination by active galactic nuclei, so the quantitative measurements remain somewhat uncertain, but it is notable that there are also indications of a similar deficiency in metals in the stellar component of some galaxies at these redshifts -- \cite{Gallazzi2014} determined that some massive star-forming galaxies have  stellar metallicities a factor of three below the solar value. More recently, work done with deep spectroscopy from the Large Early Galaxy Census (LEGA-C) survey \citep{Beverage2021} showed that some massive quiescent galaxies at a redshift of $\sim 0.7$ can have stellar iron abundances as low as [Fe/H]$\sim-0.4$.  Quite why these massive metal-deficient galaxies existed, and how they disappear before the present day, remain open questions.

To address these issues and understand the underlying physics, the chemical evolution of galaxies needs to be modeled self-consistently with their star-formation histories. A number of such general chemical evolution models have been successfully constructed \citep[e.g.][]{Spitoni2017,Lian2018mzr,Yates2021}, but they typically focus on average properties of the galaxy population, which means that they can effectively reproduce global relations like the MZR, but not the variations seen between galaxies.  Modelling of the detailed chemical evolution of individual galaxies has largely focused on the Milky Way, with its wealth of information about individual stars (as reviewed by \citealt{Matteucci2021}).  However, the quality of spectral data now becoming available for other galaxies mean that we have reached a point where ``semi-analytic spectral fitting'' allows us, at least approximately, to physically model the complete mass and metallicity history of a galaxy over its lifetime, by fitting a self-consistent evolutionary model directly to its absorption-line spectrum, while using emission-line diagnostics to constrain the present-day metallicity and star-formation rate (\citealt{Zhou2022}, hereafter Paper~I).

To-date, this technique has been applied to nearby galaxies from the SDSS-IV Mapping Nearby Galaxies at Apache Point Observatory (MaNGA; \citealt{Bundy2015}) survey, as described in Paper~I and \cite{Zhou2022environment}.  However, the same approach can also be applied to spectra of individual galaxies at intermediate redshift, to determine how they arrived at their observed masses and metallicities. Such analysis requires high-quality spectra, but such data are now available from LEGA-C, whose galaxies have a redshift of $\sim 0.7$.  We have also identified how we can  use the broad range of photometric data available for the galaxies in the LEGA-C survey to supplement the spectral constraints in this modelling process, and how to allow for dust obscuration and possible AGN contamination in the emission-line constraint on gas-phase metallicity. Moreover, we can use the similarly-reconstructed life histories of MaNGA galaxies to predict where they would lie in the mass--metallicity plane at any redshift, so we can compare them directly to the properties of LEGA-C galaxies in this plane at $z \sim 0.7$.  By matching MaNGA galaxies to their LEGA-C ``twins'', we can not only determine the evolutionary path by which the LEGA-C galaxies evolved to their observed location in the mass--metallicity plane, but also what subsequent evolution they might be expected to undergo to the present day, leading them to vacate the low-metallicity region of the plane, and what physical process drives that evolution.

Our description of this comparison is set out as follows. The data from LEGA-C and MaNGA, sample selection and data reduction are described in \S\ref{sec:data}. The fitting of these spectral data using a full chemical evolution model is discussed in \S\ref{sec:analysis}. The results and their implications are presented in \S\ref{sec:results}, and the work is summarized in \S\ref{sec:summary}. Throughout this work we use a standard $\Lambda$CDM cosmology with $\Omega_{\Lambda}=0.7$, 
$\Omega_{\rm M}=0.3$ and $H_0$=70\kms Mpc$^{-1}$.

\section{The data}
\label{sec:data}
In this section, we briefly summarize the surveys on which this analysis draws, and describe how the requisite subset of observations has been selected and processed prior to fitting to the semi-analytic galaxy evolution model.

\subsection{The surveys}
\subsubsection{LEGA-C}
\label{subsec:LEGAC}
In this work we make use of the final data release \citep{vanderWel2021} of the LEGA-C Public Spectroscopic Survey \citep{vanderWel2016,Straatman2018}. LEGA-C is a deep spectroscopic survey targeting galaxies with redshifts in the range $0.6<z<1$ in the COSMOS field \citep{Scoville2007}. The primary samples of LEGA-C are selected to satisfy a redshift-dependent $K_{\rm s}$ magnitude limit  
\begin{equation}
\label{eq:selection}
    K_{\rm s,LIM}<20.7-7.5 \log((1+z)/1.8),
\end{equation}
using the public UltraVISTA catalogue \citep{Muzzin2013}. This selection results in a sample with an overall stellar mass limit of $\sim 10^{10}{\rm M}_{\odot}$. 

Observations were carried out using VIMOS \citep{LeFevre2003} to produce high-quality spectra covering a wavelength range of 0.63--0.88$\,\mu m$ with a resolution of $\sim3500$ and average
continuum signal-to-noise ratio of $\sim$20 {\AA}$^{-1}$ \citep{vanderWel2021}.  In addition, the released LEGA-C catalogue provides data products including measurements of the emission line properties of LEGA-C galaxies, which are also used in this analysis.

As well as the spectra and related measured properties from the released LEGA-C catalogue, we also make use of the multi-band photometry for these galaxies obtained from the UltraVISTA catalogue \citep{Muzzin2013}. This catalogue provides PSF-matched photometry in 30 bands from 0.15 to 24$\,\mu m$. Data in the catalogues contain 
\begin{itemize}
    \item near-infrared data from UltraVISTA \citep{McCracken2012};
    \item optical data using broad-band $(g^+r^+i^+z^+B_jV_j)$ and 12 medium band (IA427–IA827) filters with Subaru/SuprimeCam as well as $u^* $data from CFHT/MegaCam \citep{Taniguchi2007};
    \item FUV and NUV fluxes from GALEX \citep{Martin2005};
    \item 3.6$\mu {\rm m}$, 4.5 $\mu {\rm m}$, 5.8 $\mu {\rm m}$, 8.0 $\mu {\rm m}$ and 24 $\mu {\rm m}$ data from Spitzer \citep{Sanders2007} .
\end{itemize}
Because dust emission becomes prominent in the 24\,$\mu {\rm m}$ band \citep{conroy2013}, and we do not model such emission in our fitting process, we cannot use these data, but include all the remaining 29 data points from the spectral energy distribution (SED) of each galaxy, along with the spectral data, to constrain the modelling.

\subsubsection{MaNGA}
\label{subsec:MaNGA}
The MaNGA survey \citep{Bundy2015,Yanb2016} was one of the three key programmes in SDSS-IV \citep{Blanton2017}. It successfully obtained spatially resolved spectroscopy for more than 10,000 galaxies in the local Universe ($z\sim0.03$, \citealt{Drory2015, Law2015}), with spatial 
coverage out to between 1.5 and 2.5 effective radii for most targets \citep{Wake2017}. Using the dual-channel spectrographs \citep{Smee2013} on the 2.5-meter telescope \citep{Gunn2006}, MaNGA provided high resolution (R$\sim$2000) spectra in a wavelength range from 3622{\AA} to 10354{\AA}. The raw MANGA data were processed using the bespoke MaNGA Data Reduction Pipeline \citep{Law2016}, which produces science-ready spectra with flux calibration to better than 5\% in most of the wavelength range \citep{Yana2016}. The spectra were further processed through the Data Analysis Pipeline (DAP, \citealt{Westfall2019,Belfiore2019}) to provide science data products such as the stellar and gas kinematics, emission-line fluxes and spectral indices.

\subsection{Sample selection and data preparation}
Since the modelling technique adopted here uses both absorption and emission lines to constrain the stellar and chemical evolution, we have to select sub-samples of galaxies where these properties are adequately determined, and then process them as follows.

\subsubsection{LEGA-C}
While we can accommodate a level of contamination from narrow-line AGN \citep{vanderWel2021}, strong activity that potentially distorts the whole SED would lead to spurious results in the model fitting.  Accordingly, we first exclude the 107 objects flagged in the LEGA-C data as strong AGN based on their mid-infrared or X-ray properties.

To ensure that there is sufficient signal in the continuum to constrain the properties of the stellar population, we place a cut on the spectral signal-to-noise ratio at $S/N > 20$ {\AA}$^{-1}$, using the averaged values provided by the LEGA-C catalogue. For the gas properties, we make use of the emission line strengths provided by the data release catalogue. To obtain the gas-phase metallicity of each galaxy, we make use of the calibration developed by \citealt{Kobulnicky2004} (hereafter KK04). This method uses three emission lines,
\hbox{[O\,{\sc iii}]}$\lambda$5007, \hbox{[O\,{\sc ii}]}$\lambda$3727 and  H${\beta}$. The requirement that all three lines lie within the wavelength range covered by the LEGA-C observations restricts us to the redshift range $0.6<z<0.8$.  We also require that all three lines be strong enough to measure, which imposes a limit of $S/N >5$ in their catalogued values, and we further require the equivalent width of the H$\beta$ emission line to be greater than $2\,$\AA, to ensure the galaxy has a detectable level of star-formation activity.  These criteria produce a well-defined sample of 152 galaxies at a redshift of $\sim 0.7$.

As mentioned above, these spectral data are supplemented by a large number of photometric observations. To place the spectra on the same scale as these data and account at least crudely for slit losses relative to the larger $2.1\,{\rm arcsec}$ photometric aperture, we computed synthetic photometric fluxes of the LEGA-C spectra in the subset of photometric
bands that are covered by the spectroscopy observation. The spectral fluxes were then normalized such that the fluxes in this subset of photometric bands are closest to the photometric observations in a least-squares sense \citep{Cappellari2022}. 

The next step is to derive the gas-phase metallicities of the sample galaxies, which will be used to constrain the model. As stated above, we have adopted the KK04 approach, which calibrated three emission lines against stellar evolution and photoionization grids from \cite{Kewley2002} to determine the oxygen abundance. One complication is that this calibration is bimodal depending on the ionization state of the gas. However, in the current sample, since the galaxies are all relatively massive and the redshift is not very high, the galaxies are all very likely to lie on the ``upper branch'', so we use the formula for this branch to estimate metallicity. A further issue is that because the lines used are spread over a significant range in wavelengths, the results are quite sensitive to dust attenuation.  Unfortunately, the limited spectral range of LEGA-C means that we cannot use the usual Balmer decrement approach to calibrate out such attenuation.  Instead, we adopt an iterative approach, described in detail in  \autoref{sec:fitting}, which uses the attenuation derived during the full spectral fitting process to determine the appropriate correction to the measured line ratios. Finally, we note that the KK04 calibration is sensitive to the presence of AGN contamination, and, again, the limited spectral range of the data makes it difficult to determine the level of contamination.  Accordingly, as also explained in more detail in \autoref{sec:fitting}, where there is any indication of AGN contamination, we only use a conservative lower limit for the gas-phase metallicity; it turns out that even this fairly weak limit is sufficient to constrain the ultimate fit in a useful manner. In estimating metallicity, to allow a direct comparison between gas and stellar properties, we follow paper~I by assuming a solar metallicity of 0.02 \citep{Anders1989}, consistent with the \citet{BC03} SSP models used in the spectral analysis (see \autoref{sec:analysis}). All relations are scaled to this solar abundance unless specifically stated. Although absolute values will vary if a different calibration is adopted, none of the systematic trends found in our analysis are affected.

\subsubsection{MaNGA}

For the low-redshift comparison sample, we want to start with as wide a selection as possible, because we do not know a priori how the LEGA-C galaxies will evolve.  Accordingly, we adopt the large sample used in  \cite{Zhou2022}. This selection is drawn from the final full survey made available in the 11th MaNGA Product Launch data, as part of SDSS-IV Data Release 17 (DR17, \citealt{SDSSDR17}). From this full data set, all the reasonably face-on ($\frac{b}{a}>0.5$) galaxies were selected.  Since we are seeking to compare global properties, all the data from within the central effective radius were co-added to produce a single high signal-to-noise ratio spectrum for analysis; any with $S/N<30$ were excluded. Finally, we selected galaxies detected with $S/N > 5$ in four emission lines, \hbox{[O\,{\sc iii}]}$\lambda$5007, \hbox{[N\,{\sc ii}]}$\lambda$6584, H${\alpha}$, and H${\beta}$, to provide the gas-phase constraints on metallicity and star-formation rate.  This selection yielded a large final sample of 2560 objects as potential descendants of the LEGA-C galaxies.

Note that the wide wavelength coverage of the MaNGA data allows us to use alternative diagnostics of the gas-phase metallicity. In particular, we can make use of the O3N2 calibration (\citealt{PP04}, hereafter PP04), which uses line ratios of pairs closely spaced in wavelength, so is less sensitive to the effects of dust attenuation than the KK04 calibration.  As also noted by \cite{Kumari2019}, O3N2 is also less affected by AGN. We will use the fact that we can compare these two measures of metallicity in the MaNGA spectra to inform how we use the KK04 measure in the LEGA-C data (see \autoref{sec:analysis} and the Appendix).  As noted by  \cite{Kewley2008}, the KK04 calibration has a systematic offset compared to the PP04 calibration. To ensure that the two measures remain consistent, we make use of the conversion polynomial provided by \cite{Kewley2008} to convert oxygen abundance obtained using KK04 to values similar to the PP04 O3N2 approach. In addition, in order to compare gas and stellar metallicities, we assume a solar abundance pattern to convert the derived oxygen abundance to metallicity in solar units.

\section{Analysis}
\label{sec:analysis}

\subsection{The chemical evolution model}
\label{sec:model_legac}
To fit the mass and chemical evolution of LEGA-C galaxies, we adopt a fairly simple chemical evolution model similar to Paper~I; a detailed description and extensive tests of the model are presented there.  In brief, this model seeks to embrace the main processes of star formation, gas inflow and outflow to characterise the chemical evolution of a galaxy's gas reservoir. In this model, the mass evolution of the gas in a galaxy can be described by the equation
\begin{equation}
\label{eq:massevo}
\dot{M}_{\rm g}(t)=
   A e^{-(t-t_0)/\tau}-S(1-R)M_{\rm g}(t)-S\lambda M_{\rm g}(t).
\end{equation}
The first term characterises the gas infall into the galaxy, quantified by a commonly-adopted exponentially decaying form specified by two physical parameters, a starting time $t_0$ and a timescale of the infall $\tau$. The second term takes into account the star formation process that removes gas from this reservoir. We assume a linear Schmidt law, with star formation rate $\psi(t)=S\times M_{\rm g}(t)$
\citep{Schmidt1959} and a star formation efficiency $S$
estimated from the stellar mass surface density ($\Sigma_{*}$) via the formula proposed by \cite{Shi2011},
\begin{equation}
    S (yr^{-1})=10^{-10.28\pm0.08} \left( \frac{\Sigma_*}{M_{\odot} pc^{-2}} \right)^{0.48}.
\end{equation}
The mass return fraction $R$ accounts for the return of gas from dying stars, which we set to $R = 0.3$ for all generations.
The final term in \autoref{eq:massevo} describes outflow from the galaxy. We adopt the conventional assumption that the outflow strength is proportional to the SFR, with a ‘wind parameter’ $\lambda$ quantifying its relative strength \citep{Arimoto1987}.

To characterise the metallicity evolution of this gas component, we adopt the instantaneous mixing approximation\footnote{More properly, one should follow the timescales over which elements are returned to the galaxy from stars of different types. We plan to explore modeling this process more realistically in future.}, so that the gas is assumed to be always well mixed during its evolution. Taking all the physical processes into account, the equation that characterizes the chemical evolution of the gas can then be written
\begin{equation}
\label{eq:cheevo1}
\begin{aligned}
\dot{M}_{Z}(t)=& Z_{\rm in}\dot{M}_{\rm in}(t)-Z_{\rm g}(t)(1-R)SM_{\rm g}(t) +\\
& y_Z(1-R)SM_{\rm g}(t) -Z_{\rm g}(t)\lambda SM_{\rm g}(t),
\end{aligned}
\end{equation}
where $Z_{\rm g}(t)$ is the gas phase metallicity and $M_{Z}(t)\equiv M_{\rm g}\times Z_{\rm g}$ is the total mass in metals. The first term on the right hand side of the equation is the inflow term, which brings gas with metallicity $Z_{\rm in}$ into the system; we assume the infall gas is pristine so that $Z_{\rm in}=0$.  The second term describes the mass of metals locked up in long-lived stars. The third term characterises the return of chemically-enriched gas from dying massive stars, with the free parameter $y_Z$  being the fraction of metal mass generated per stellar mass. The last term represents the removal of metal-enriched gas by the outflow.

\subsection{Spectral fitting of LEGA-C data}
\label{sec:fitting}

\begin{table*}
	\centering
	\caption{Priors of model parameters used to fit the LEGA-C data}
	\label{tab:paras}
	\begin{tabular}{lccr}
		\hline
		Parameter & Description & Prior range\\
		\hline
		$y_Z$ & Yield parameter & $[0.0, 0.08]$\\
		$\tau$ & Gas infall timescale & $[0.0, 10.0]$Gyr\\
		$t_{0}$ & Start time of gas infall & $[0.0, 8.0]$Gyr\\
		$\lambda$ & The wind parameter & $[0.0, 10.0]$\\
		$\tau_{\rm V, Y}$ & Dust optical depths for the young (<100 Myr) stellar population  & $[0.0, 2.0]$\\   
		$\tau_{\rm V, O}$ & Dust optical depths for the old (>100 Myr) stellar population  & $[0.0, 1.0]$\\
		\hline
	\end{tabular}
\end{table*}

We vary the parameters of this model to find the best fit to the spectral and photometric data for each LEGA-C galaxy, using a Bayesian approach. In brief, a set of model parameters 
(i.e. $y_Z,t_0,\tau,\lambda$ for chemical revolution modelling, and the dust attenuation parameters described below) are generated from a prior distribution listed in \autoref{tab:paras}. Note that the starting time of the gas infall is constrained by the age of the Universe at $z \sim 0.7$ to be more recent than $8\,{\rm Gyr}$ ago. These parameters are then used to calculate the star-formation history (SFH) and chemical evolution history (ChEH) following \autoref{eq:massevo} and \autoref{eq:cheevo1}. The SFH and ChEH are combined with the \cite{BC03} SSP models to calculate a composite stellar population following the usual stellar population synthesis approach (see \citealt{conroy2013} for a review). 
At optical wavelengths, we use 
\cite{BC03} models constructed using the Chabrier IMF \citep{Chabrier2003}  and the STILIB empirical stellar spectra templates \citep{Borgne2003}. These SSPs cover metallicities from Z = 0.0001 to Z = 0.05, and ages from
0.0001 Gyr to 20 Gyr, with a resolution
of 3{\AA} in a wavelength range 3200–-9500{\AA}. For dust attenuation, we use the two-component \cite{Charlot2000} model to assign different dust optical depths, $\tau_{\rm V,Y}$ and $\tau_{\rm V,O}$, to young (<100 Myr) and old (>100 Myr) stellar populations. In addition, to make use of the information contained in the full SED of the galaxy, we use \cite{BC03} models with the same IMF and isochrone settings, but constructed using the low-resolution Basel theoretical templates that cover the full spectral range from 91{\AA} to 160 $\mu$m. After redshifting to the observed frame and multiplying by each filter's response function, we obtain a synthetic SED for the model that can be directly compared with the observations. 

The derived spectrum, gas-phase metallicity and SED are compared with the observed data using a $\chi^2$-like likelihood function,
\begin{equation}
\label{likelyhood:legac}
\ln {L(\theta)}\propto -
\frac{(Z_{\rm g,\theta}-Z_{\rm g,D})^2}{2\sigma_{Z}^2
}-\sum_{j=1}^M\frac{(F_{\theta,j}-F_{\rm D,j})^2}{2F_{err,j}^2}-\sum_{i=1}^N\frac{\left(f_{\theta,i}-f_{\rm D,i}\right)^2}{2f_{\rm err,i}^2},
\end{equation}
where $f_{\theta, i}$ and $f_{\rm D, i}$ are the model and observed flux at the $i$-th wavelength point in the spectrum, with 
$f_{\rm err,i}$ being the corresponding error and $N$ the total number of wavelength points.  Similarly, $Z_{\rm g,\theta}$ and $Z_{\rm g,D}$ are the current gas-phase metallicities predicted from the chemical evolution models and from the observed data respectively, and $\sigma_{Z}$ is the estimated uncertainty, and $F_{\theta, j}$ and $F_{\rm D, j}$ are the model and observed flux in the $j$-th photometric band, with $F_{\rm err,j}$ being the corresponding error and $M=29$ the total number of bands.

As discussed in \autoref{sec:data}, the gas-phase metallicities used in the constraints in \autoref{likelyhood:legac} are derived using the KK04 calibration, in which the line ratios used in the calculation can be affected significantly by dust attenuation in the galaxy, whose value we do not initially know. To account for this effect, we perform the spectral fitting in an iterative manner.  We first derive an estimate of the gas-phase metallicity of a galaxy directly from the observed line flux using the KK04 calibration, and carry out the spectral fitting procedure using this estimate neglecting attenuation to constrain the present-day metallicity of forming stars.  As described above, the model contains two components of dust attenuation, and the value associated with young stars should correspond closely to the attenuation experienced by light from the gas, since the stars are forming directly from this material. Accordingly, we use the derived attenuation of the young stellar component as an estimate for the attenuation of light from the gas, correct the line strengths appropriately, update the derived metallicity estimate, and redo the fit.  We repeat this process until it converges, which we take to be the gas-phase metallicity varying by less than 5\% between iterations; typically, this procedure converges within five iterations. At this point, we have both the final estimate of the gas-phase metallicity and the best-fit evolutionary model for the galaxy.

One remaining issue is that, as previously noted, the KK04 calibration is also sensitive to contamination from AGN. To assess their possible impact, we have used the MaNGA sample, for which we can also use the O3N2 metallicity indicator, which is relatively immune to the effects of the narrow-line AGN that remain in the sample \citep{Kumari2019}. As we show in the Appendix \ref{appedix_caliration}, when the gas phase metallicities are relatively high ($\log(Z_{\rm g}/Z_{\odot})>-0.2$), the KK04 calibrator remains reliable, but in the low metallicity regime ($\log(Z_{\rm g}/Z_{\odot})\lesssim-0.2$) there is a significant bias due to the AGN contamination.  Accordingly, if the gas-phase metallicity we have derived iteratively for a LEGA-C galaxy has $\log(Z_{\rm g}/Z_{\odot})>-0.2$, we accept the value and the corresponding best-fit evolutionary model for the galaxy.  If, however, we find a value of $\log(Z_{\rm g}/Z_{\odot})<-0.2$, we are in the regime where AGN activity may have biased the result. In this case, we treat the value obtained from the KK04 calibration, denoted $Z_{\rm g, KK04}$, as a conservative lower limit on the true gas-phase metallicity.  We thus refit the galaxy evolutionary model to the spectral and photometric data using a modified likelihood function, 
\begin{equation}
\label{eq:like_alter}
\begin{aligned}
\ln {L(\theta)}\propto
&
 -\left\{\begin{array}{lr}
   \frac{(Z_{\rm g,\theta}-Z_{\rm g, KK04})^2}{2\sigma_{Z}^2} &\text{(for $Z_{\rm g,\theta}<Z_{\rm g, KK04}$)}\\
   0 &\text{(for $Z_{\rm g,\theta}>Z_{\rm g, KK04}$)}
   \end{array}.
   \right .
\\
&
-\sum_{i}^N\frac{\left(f_{\theta,i}-f_{\rm D,i}\right)^2}{2f_{\rm err,i}^2}
-\sum_{j}^M\frac{(F_{\theta,j}-F_{\rm D,j})^2}{2F_{err,j}^2}
\end{aligned}
\end{equation}
During the fitting process with such a likelihood function, when the best-fit gas-phase metallicities do not converge towards the boundary set by the KK04 calibration, the effective constraints come from the optical spectra and the SED.

In the tests presented in Paper~I, we showed that this method recovers the simulated star formation and chemical evolution history that we input, as long as we have data of the quality of MaNGA spectra. In this work, we have the smaller spectral range of the LEGA-C data to consider, and can lack a reliable estimate of gas-phase metallicity, but have the supplementary information provided by the full SED. It is therefore worth repeating some tests to see how these differences affect the reliability of the results when using the LEGA-C data. As shown in Appendix~\ref{appedix_mocktest}, the inclusion of the constraints from the SED goes a long way toward recovering the information that we lose in the absence of a measure of current gas-phase metallicity, allowing reasonably robust estimates of both star formation history and metallicity.

\begin{figure*}
    \centering
    \includegraphics[width=1.0\textwidth]{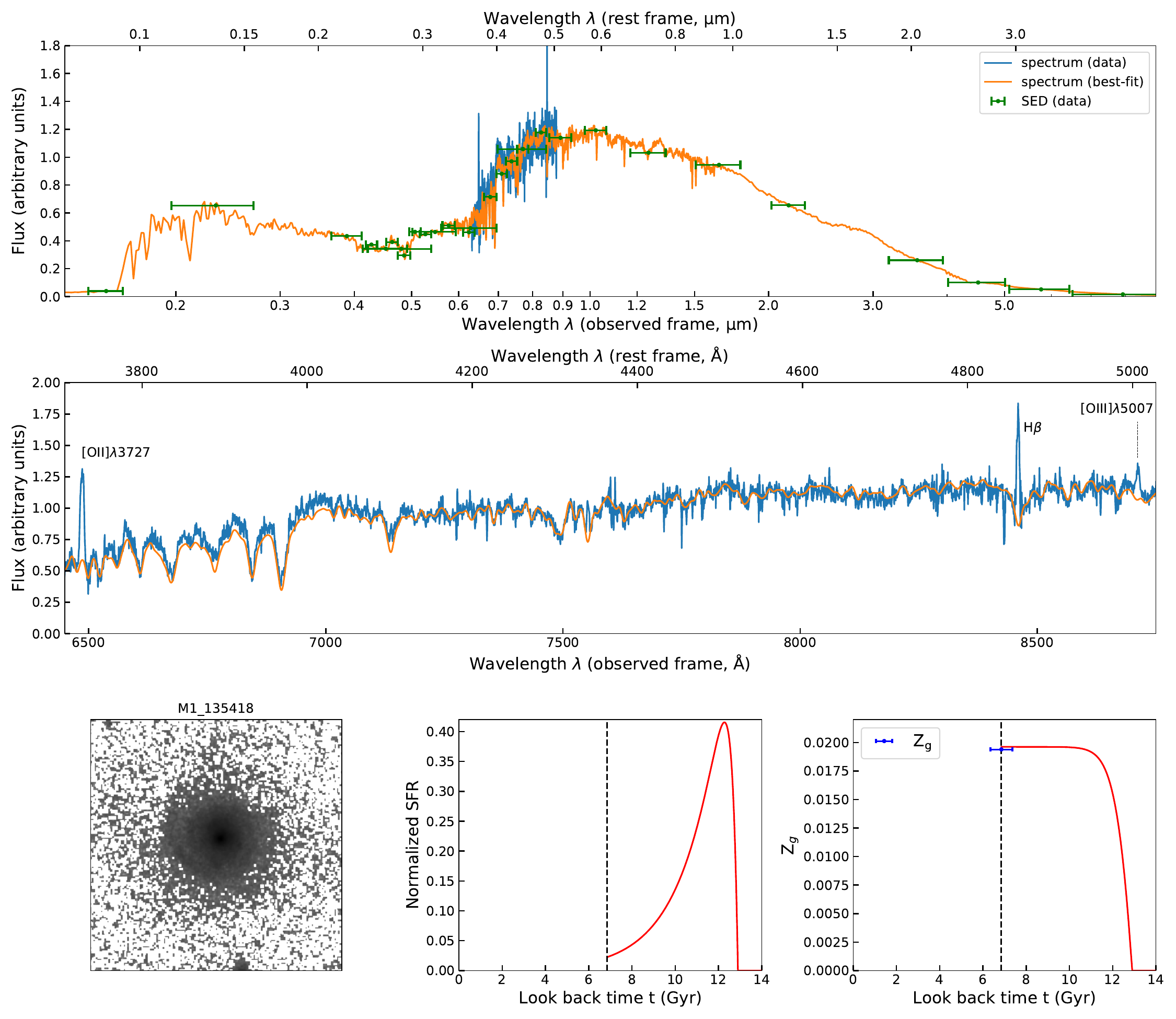}\\
     \caption{An example showing the fit to the spectrum and SED of a LEGA-C galaxy. The top panel shows the entire wavelength range covered by the SED adopted in this work, while the panel in the middle zooms in to the region covered by the LEGA-C spectral data, with relevant emission lines labelled. The blue line shows the spectral data while the green horizontal bars are the photometric measurements and their bandpasses. The orange line is the best fit to these data. The bottom-left panel shows the HST-ACS image of the galaxy. The middle and right panels in the bottom row show the star-formation history and chemical evolution of the gas phase for the best-fit model up to the observed redshift (marked by a vertical dashed line). The small blue bar in the bottom right panel indicates the gas-phase metallicity at the observed redshift as inferred directly from emission lines.}
     \label{fig:exmaple_fitting}
\end{figure*}

An example of the final output from this process is presented in \autoref{fig:exmaple_fitting}.  It can be seen that the fit can simultaneously reproduce the spectral data, the broader SED, and the constraint from emission lines on metallicity, returning an entirely reasonable self-consistent evolutionary history in both star formation and metallicity evolution. We have carried out similar fits for the data from all 152 LEGA-C galaxies in the sample.

\begin{figure}
    \centering
    \includegraphics[width=0.5\textwidth]{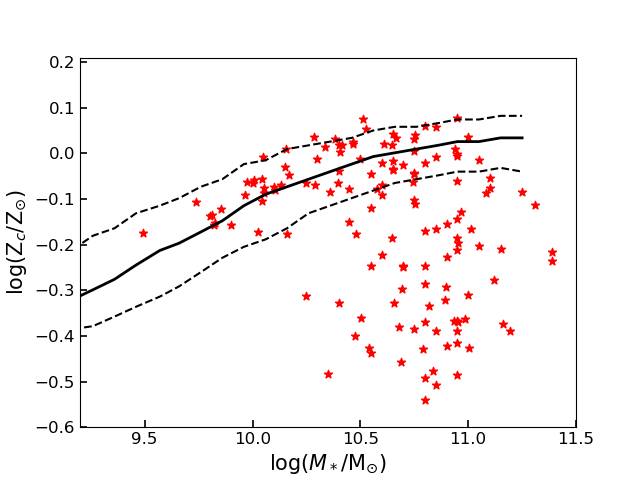}\\
     \caption{The distribution of LEGA-C galaxies in the mass--metallicity plane.  Both the stellar mass $M_*$ and the current gas-phase metallicity $Z_c$ are the values inferred from the best-fit model for each galaxy
 The black solid line shows the local mass--metallicity relations from \citealt{Tremonti2004}, with dashed lines indicating the 16th and 84th percentiles of the spread in the local sample.}
     \label{fig:mzr_LEGAC}
\end{figure}

One benefit of this fit is that we now have an estimate of the current gas-phase metallicity that is robust against the influence of AGN, since we can use the value from the model fit to the stellar population -- in essence, we are using the metallicity of the youngest stars in the composite absorption-line spectrum being fitted rather than the potentially-biased value from emission lines in the gas from which those stars formed.  To avoid confusion with the less-reliable direct estimates of gas-phase metallicity, we refer to this ``current'' metallicity as $Z_{\rm c}$.  We can therefore now produce a view of the mass--metallicity plane at the $\sim 0.7$ redshift of the LEGA-C sample by plotting $Z_{\rm c}$ against the stellar mass of the best-fit model for all the galaxies in the LEGA-C sample, as shown in \autoref{fig:mzr_LEGAC}.

For reference, we also show in the figure the local gas-phase MZR derived by \cite{Tremonti2004}, converted to the adopted PP04 calibration and into solar units.  It is apparent that a good number of the LEGA-C galaxies already lie close to the relation, so seemingly have largely completed their evolution in the mass--metallicity plane. However, we also find the secondary population noted in the Introduction of high-mass low-metallicity galaxies that are not seen in the nearby Universe. Such a distribution of galaxies on the mass--metallicity plane is closely consistent with what \cite{Lewis2023} have recently found, also using the LEGA-C data set. The main difference from the \cite{Lewis2023} result is that we have deliberately excluded most metal-poor galaxies with relatively low stellar masses from our sample because we require a high signal-to-noise ratio in the stellar continuum of the spectral data in order to determine the stellar properties of the galaxies. In addition, \cite{Lewis2023} use the MEx diagram \citep{Juneau2011} to exclude galaxies that potentially have AGN contamination, which may also exclude some massive metal-poor galaxies (see discussion in \autoref{sec:origins}). With the full evolutionary models that we have fitted to these data, we are now in a position to begin to look at the physical processes that might have led them to this location in the mass--metallicity plane, and, perhaps more importantly, by matching them to the properties of the nearby MaNGA galaxies, we may be able to see why they have all subsequently disappeared.

\subsection{Spectral fitting of MaNGA data}
\label{sec:analysis_MAGNA}

With the analysis of the LEGA-C galaxies in hand, we seek to identify their descendants in the local Universe. To this end, we investigate the evolution of MaNGA galaxies by analysing their integral spectra to reconstruct their formation histories and tracing them back to $z\sim0.7$. The semi-analytic spectral modelling approach is the same as for the LEGA-C data, but with one additional component.  Our analysis in Paper~I found that at low redshifts an additional ingredient is required for the model -- a cut-off in the wind-driven mass-loss.  Such a phenomenon might be expected for a variety of physical reasons: it could represent the gravitating mass of the galaxy growing to the point where stellar processes becoming ineffective at expelling material from the system, or it could be that declining black hole activity leads to AGN ceasing to be a mechanism by which material is blown out of galaxies. Whatever its cause, it can be incorporated into the simple parametric model we are using to characterise galaxy evolution by simply shutting off the wind at some time $t_{\rm cut}$, so that \autoref{eq:massevo} becomes
\begin{equation}
\label{eq:massevo_MANGA}
\dot{M}_{\rm g}(t)=
 \left\{\begin{array}{lr}
   A e^{-(t-t_0)/\tau}-S(1-R)M_{\rm g}(t)-S\lambda M_{\rm g}(t) &\text{(for $t<t_{\rm cut}$)}\\
   A e^{-(t-t_0)/\tau}-S(1-R) M_{\rm g}(t) &\text{(for $t>t_{\rm cut}$)}
   \end{array}.
   \right.
\end{equation}
Accordingly, the chemical evolution can be written 
\begin{equation}
\label{eq:cheevo1_MANGA}
\begin{aligned}
\dot{M}_{Z}(t)=& Z_{\rm in}Ae^{-(t-t_0)/\tau}+S(1-R)(y_Z-Z_{\rm g}(t)) M_{\rm g}(t)\\
& -Z_{\rm g}(t)\lambda SM_{\rm g}(t),
\end{aligned}
\end{equation}
 and 
\begin{equation}
\label{eq:cheevo2_MANGA}
\dot{M}_{Z}(t)=Z_{\rm in}Ae^{-(t-t_0)/\tau}+S(1-R)(y_Z-Z_{\rm g}(t)) M_{\rm g}(t),
\end{equation}
for $t<t_{\rm cut}$ and  $t>t_{\rm cut}$, respectively.

The spectral fitting procedure is then very similar to that applied to the LEGA-C galaxies, but with one more free parameter, $t_{\rm cut}$. The only other differences are that we are able to use gas-phase metallicities derived using the more robust PP04 approach to constrain the present-day metallicity of the galaxies, and that we can use a simpler screen dust model specified with a \cite{Calzetti2000} attenuation curve to account for dust attenuation, since we no longer need the multi-component model to iteratively correct the emission-line diagnostic. Again, more complete details of the fitting and tests of it can be found in Paper~I.  

\begin{figure}
    \centering
    \includegraphics[width=0.5\textwidth]{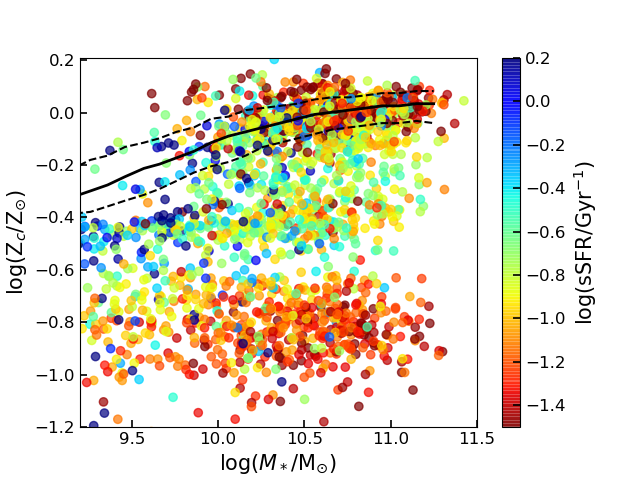}\\
     \includegraphics[width=0.5\textwidth]{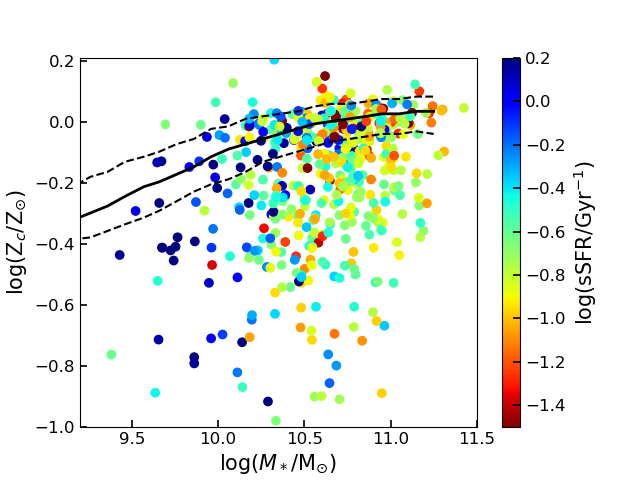}\\   
     \caption{Predicted gas-phase metallicities ($Z_{\rm c}$) as a function of the stellar masses ($M_*$) inferred from the model fits to MaNGA galaxies, as they would have been observed at the redshift of LEGA-C galaxies. Galaxies are colour-coded by their specific star-formation rates at the LEGA-C redshifts.  The top panel shows the full MaNGA data set, while the bottom panel shows the sample after applying the LEGA-C selection criteria. The black solid line shows the local mass--metallicity relation from \citealt{Tremonti2004}, with dashed lines showing the 16th and 84th percentiles of the local sample. 
     }
     \label{fig:mock_MZR_MaNGA}
\end{figure}

Having fitted all 2560 galaxies in the MaNGA sample in this way, we are able to trace their properties back to any given redshift. To generate a comparison set for the current LEGA-C sample, we randomly assigned the redshift of a LEGA-C galaxy to each MaNGA galaxy to construct a mock sample with the same redshift distribution (since the distribution is quite narrow around $z \sim 0.7$, the exact choice of allocations made little difference to the results).  For each MaNGA galaxy, we then looked at its best-fit evolutionary model to predicted the properties of mass and gas-phase metallicity at its assigned LEGA-C redshift.  The resulting distribution in the mass--metallicity plane is shown in the top panel of  \autoref{fig:mock_MZR_MaNGA}. Again, we use the notation $Z_{\rm c}$ to indicate that the gas-phase metallicities are obtained from the best-fit models rather than a direct measurement of emission lines.  While there is a well-established mass--metallicity relation in this figure, similar to that seen in the LEGA-C data, there are also many more galaxies that do not follow the zero-redshift relation than are seen in the LEGA-C sample.

However, this is not really a fair comparison, since the LEGA-C galaxies have been selected using criteria that could well make some of the MaNGA objects unobservable at this redshift.  For example, it can be seen that many of the low-metallicity galaxies in this plot have very low specific star-formation rates, so would be unlikely to show the emission lines that would allow them to be included in our LEGA-C sample.  In addition, the LEGA-C selection process has a bias toward more luminous galaxies at higher redshift \citep{vanderWel2021}, which has not been applied to the MaNGA data.

We can at least approximate these criteria for the MaNGA galaxies to create a sample that should be better matched to the LEGA-C data.  As a starting point, we use the best-fit model to predict the flux from the galaxy at its assigned redshift in the observed $K_{\rm s}$ band, and then use the LEGA-C selection criterion of \autoref{eq:selection} to decide whether the galaxy should remain in the mock sample.  The other major selection effect in the LEGA-C data is that we have required three emission lines, \hbox{[O\,{\sc iii}]}$\lambda$5007, \hbox{[O\,{\sc ii}]}$\lambda$3727 and  H${\beta}$, to be observed and strong enough to make an estimate of metallicity.  By construction, the mock MaNGA sample has been placed at redshifts where all three lines lie in the wavelength range observed in the LEGA-C spectra. The expected strengths of all the emission lines are harder to predict for the redshifted MaNGA sample since they are not something that the simple chemical evolution model specifies. Fortunately there is a simplification that we can make -- inspection of the LEGA-C data show that all the galaxies with all three lines at the appropriate strength meet the requirement on the H$\beta$ line that EW(H${\beta})>2$\AA\ and f(H${\beta})>10^{-17}\,$erg/s/cm$^2$.  Since the strength of the H$\beta$ emission line is related to star formation rate, which is a quantity predicted by the model, we can at least approximately apply a selection criterion that matches the requirement on line strengths.  

We estimate these H${\beta}$ fluxes and equivalent widths of each redshifted MaNGA galaxy as follows. We first calculate the composite spectrum from the reconstructed SFH and ChEH for the galaxy at its assigned redshift.  The continuum level around H$\beta$ is obtained using a window of 100{\AA} centred at the line's rest-frame wavelength.  We then use the Kennicutt-Schmidt law to convert the model star formation rate of the galaxy at this redshift to the H$\alpha$ line flux. The H${\beta}$ line flux is then calculated by assuming an intrinsic ratio of 2.8 between H$\alpha$ and H${\beta}$ fluxes, with a further correction for extinction assuming 1 magnitude at the H$\alpha$ line \citep{Rodriguez2017} and a Calzetti attenuation curve. For the continuum, we assume the attenuation in the stellar phase is approximately 0.44 times the attenuation in the gas phase, i.e.  $A_*\sim 0.44A_{\rm gas}$ \citep{Li2021}. The equivalent width of the H${\beta}$ line is then calculated from the dust-reddened stellar continuum level and H${\beta}$ line fluxes. 

After applying these approximate selection criteria in magnitude and emission-line strength, the redshifted MaNGA sample is reduced to  511 objects, whose position in the mass--metallicity plane is show in the lower panel of \autoref{fig:mock_MZR_MaNGA}.  Unsurprisingly, this selection has removed many of the low-metallicity galaxies from the plane, since their very low star-formation rates meant that they were unlikely to contain the emission lines necessary to be in the current LEGA-C sample.  Indeed, comparing this plot to \autoref{fig:mzr_LEGAC}, it is clear that, to within the approximations required to mimic the LEGA-C selection criteria, the redshifted MaNGA galaxies now reproduce the LEGA-C data rather well.  In what follows, we will use this matched sub-sample of redshifted MaNGA galaxies to compare with the LEGA-C data.

\section{Result and discussion }
\label{sec:results}
\subsection{The mass-metallicity relation at $z\sim0.7$}

\begin{figure}
    \centering
    \includegraphics[width=0.5\textwidth]{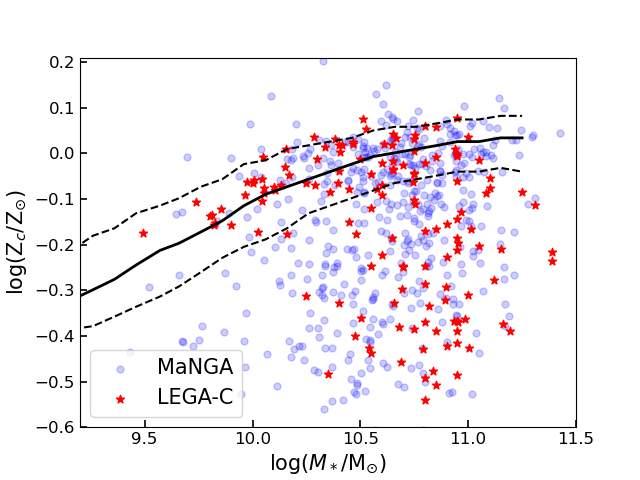}\\
     \caption{Comparison between the distribution in mass and metallicity of the LEGA-C galaxies and the values inferred for MaNGA galaxies at the redshifts of the LEGA-C data, with the LEGA-C selection criteria applied. The black solid line shows the local mass--metallicity relation from \citealt{Tremonti2004}, with dashed lines showing the 16th and 84th percentiles of the local sample. 
     }
     \label{fig:mock_MZR_MaNGA_compare}
\end{figure}

To facilitate a comparison, \autoref{fig:mock_MZR_MaNGA_compare} shows the positions of both the LEGA-C galaxies and the matched redshifted MaNGA sample in the mass--metallicity plane. From this plot, we can see that both the LEGA-C data and the redshifted MaNGA galaxies are telling the same story -- while many galaxies are already settled on the present-day mass--metallicity relation, there remains a significant population of high-mass low-metallicity galaxies that are still to reach it even at this relatively low redshift. Indications of metallicity deficiency in some massive galaxies have also found in previous investigations of galaxies at around $z\sim0.7$ \citep{Zahid2011,Maier2015,Gallazzi2014,Beverage2021,Lewis2023}, although with varying strength and relative fraction due to differences in samples and selection effects. The physical modelling used here means that we should be in a position to start to understand the processes that drive this evolution.

Indeed, the similarity of the two distributions in \autoref{fig:mock_MZR_MaNGA_compare} offers a reassuring check that the semi-analytic spectral fitting is capturing at least some of the essence of how these galaxies have evolved -- there was nothing in the fitting process applied to the MaNGA galaxies that forced them to match the properties of the LEGA-C galaxies when looking back in the derived history to the LEGA-C redshifts.  As a further check of the veracity of the reconstructed histories, we can check whether LEGA-C and MaNGA galaxies that lie in similar parts of \autoref{fig:mock_MZR_MaNGA_compare} also had similar evolutionary histories at higher redshifts that got them to this point.  Once such a similarity is established, we can with some confidence describe the LEGA-C galaxies as the likely progenitors of the current-day MaNGA galaxies, and compare other properties such as their morphologies to see how these have evolved in individual galaxies between a redshift of $\sim 0.7$ and the present day.

\begin{figure*}
    \centering
    \includegraphics[width=1.0\textwidth]{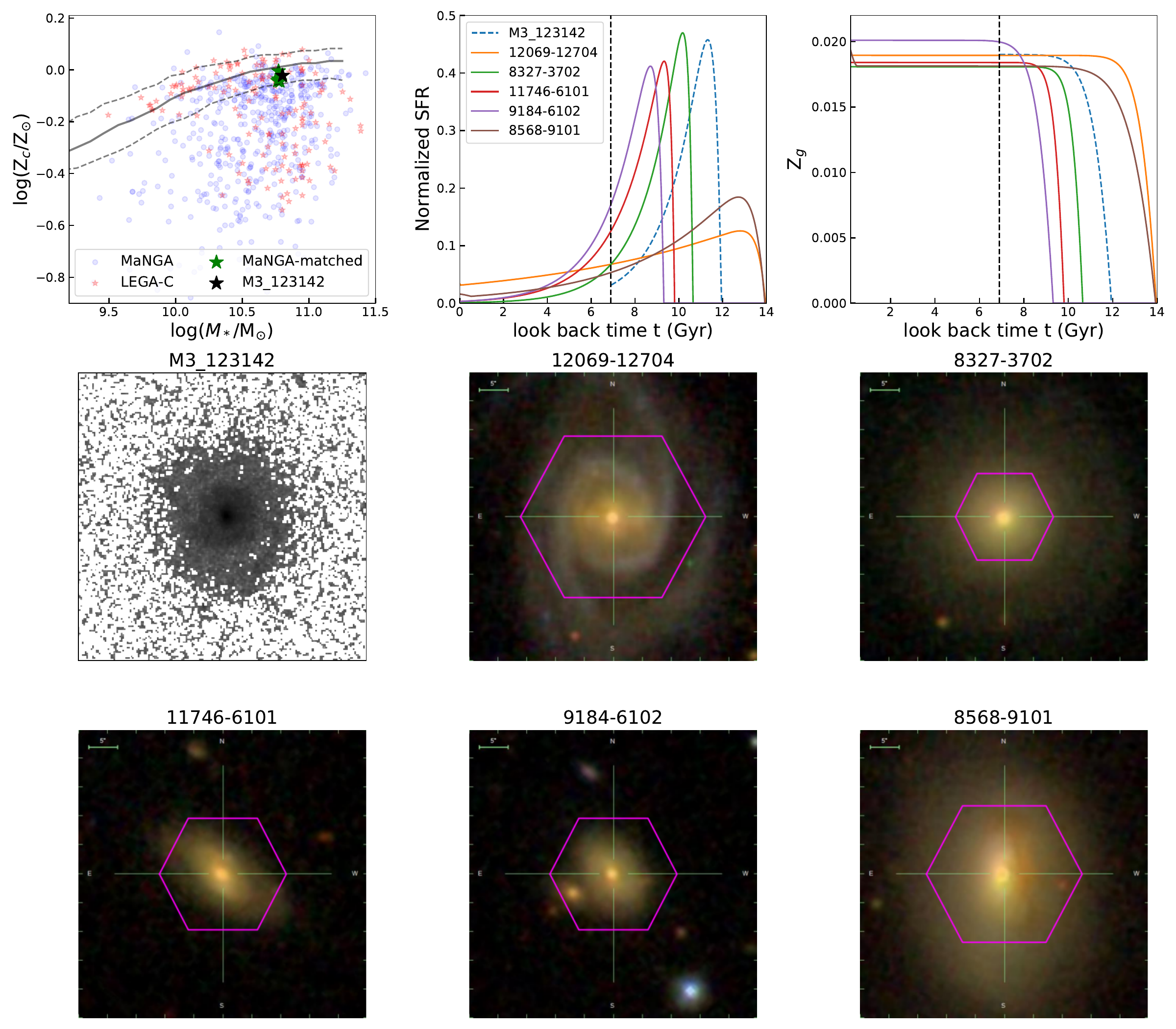}\\
     \caption{An example LEGA-C galaxy with its 5 matched MaNGA galaxies. The top left panel shows the position of these galaxies on the mass--metallicity plane. The middle and right panels in the top row show the star-formation histories and chemical evolution of these galaxies reconstructed from the model fits, with the id of the LEGA-C galaxy and \texttt{plateifu} of MANGA galaxies indicated. In each panel, the result for the LEGA-C example is shown as a blue dashed line, while the MaNGA galaxies are shown as different colour solid lines. A black vertical dash line indicates the lookback time at the observed redshift of the LEGA-C galaxy. The left panel in the second row shows the HST image of the LEGA-C galaxy, while the other 5 panels show the optical images of the matched MaNGA galaxies. 
     }
     \label{fig:example_massivehigh}
\end{figure*}

\begin{figure*}
    \centering
    \includegraphics[width=1.0\textwidth]{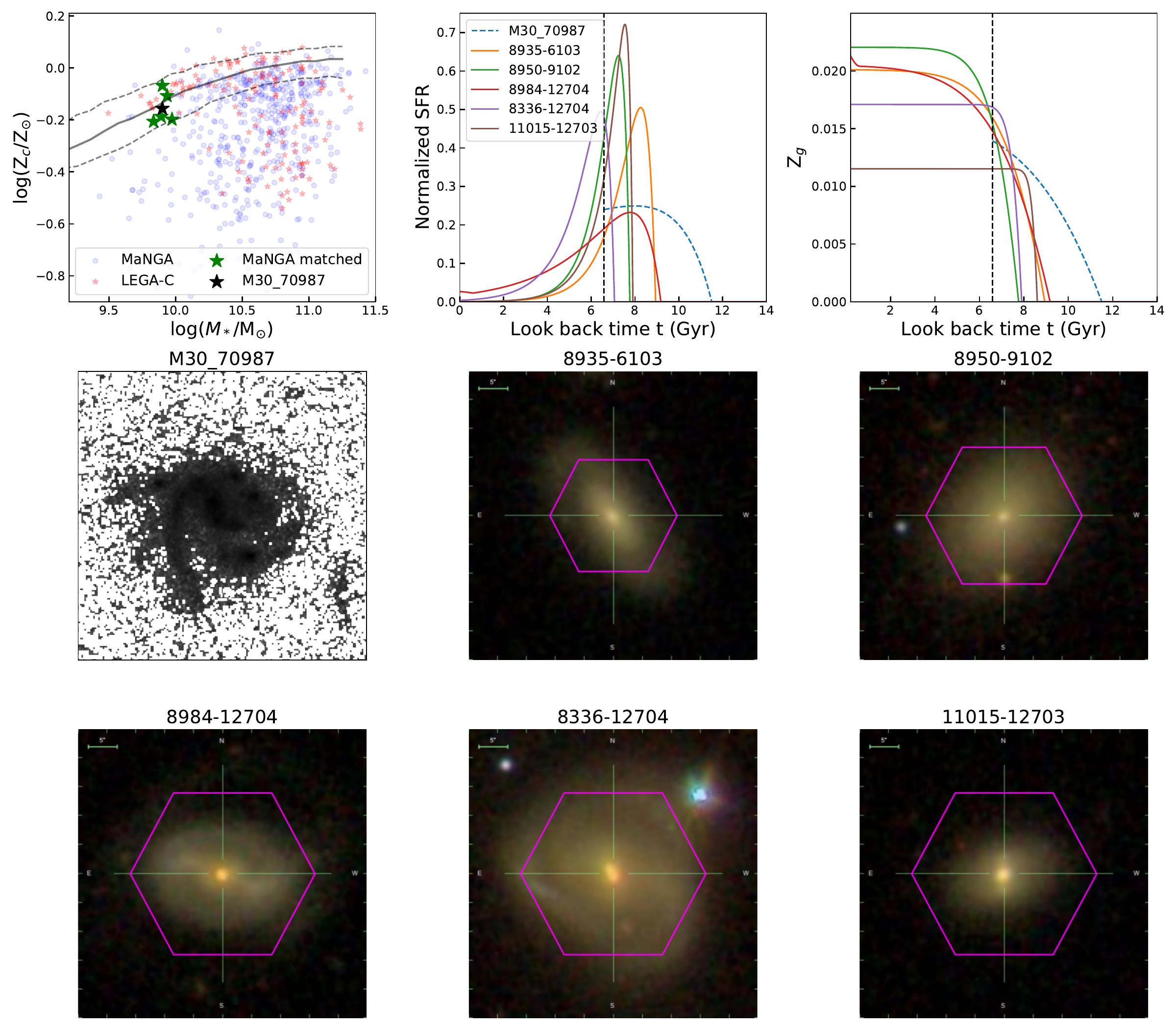}\\
     \caption{As for \autoref{fig:example_massivehigh}, but for a less massive LEGA-C galaxy. 
     }
     \label{fig:example_lowmass}
\end{figure*}

\begin{figure*}
    \centering
    \includegraphics[width=1.0\textwidth]{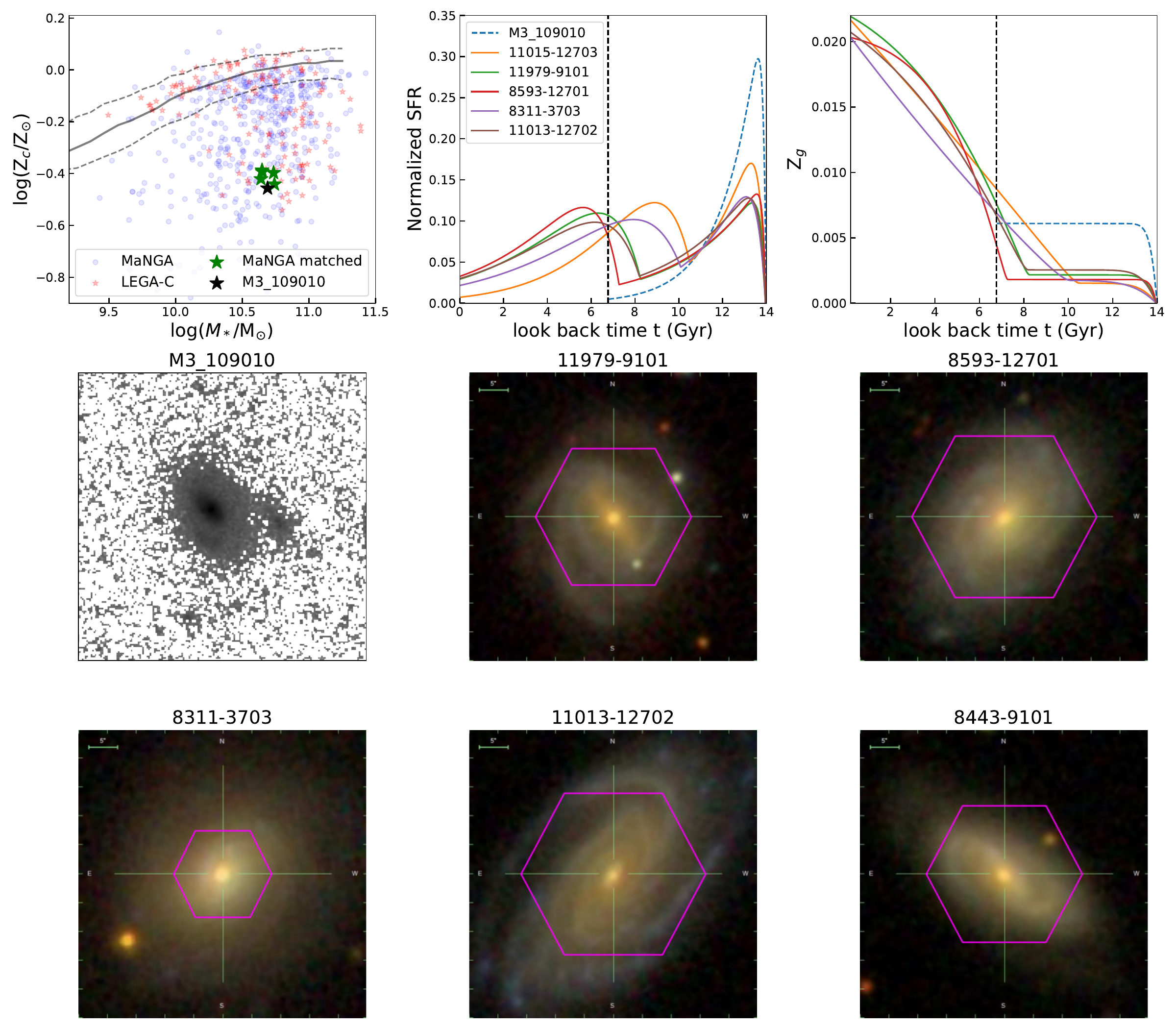}\\
     \caption{As for \autoref{fig:example_massivehigh}, but for a metal-poor LEGA-C galaxy. 
     }
     \label{fig:example_massivelow}
\end{figure*}

\subsection{Matching individual galaxies}
In order to match up galaxies between the two samples, we define a simple ``distance'' between objects in the samples in the mass--metallicity plane as  
\begin{equation}
    D\equiv\sqrt{(\log Z_{\rm c,L} - \log Z_{\rm c,M})^2+(\log M_{\rm *,L}- \log M_{\rm *,M})^2},
\end{equation}
where $Z_{\rm c}$ and $M_{*}$ are the metallicity and stellar mass in the two samples at a redshift of $\sim 0.7$. For the MaNGA galaxies, stellar masses at $z\sim0.7$ were inferred from the SFHs to that point, derived using the method described in \autoref{sec:analysis_MAGNA}. As the MaNGA comparison sample is significantly larger than the LEGA-C data set, we match each LEGA-C galaxy with its five closest potential MaNGA descendants, and compare their properties.

We start by looking at a typical example of a massive metal-rich LEGA-C galaxy, which already lies on the present-day MZR.  As \autoref{fig:example_massivehigh} shows, the model fit to this object's star-formation and chemical evolution history up to its observed epoch is indistinguishable from the MaNGA galaxies to which it has been matched, with much of the star formation and associated chemical enrichment happening at very early times.
Examining the best-fit model parameters for the galaxy, we find only a mild outflow with wind parameter $\lambda=2.2$ occurring in the galaxy, keeping it in an equilibrium state with little metallicity evolution over the last $\sim10\,{\rm Gyr}$. The subsequent evolution in the candidate MaNGA descendants is rather modest, explaining how they stay on the MZR, and why their present-day appearance in the images shown in \autoref{fig:example_massivehigh} are not that different from the LEGA-C galaxy. 

Next, we move towards the less massive end of the present-day MZR. As \autoref{fig:example_lowmass} shows, 
this sample LEGA-C galaxy has a lower metallicity than  the more massive galaxies in \autoref{fig:example_massivehigh}, which results from the stronger outflow
with $\lambda=3.4$ found in the best-fit model. There is again a reasonable match between the LEGA-C and MaNGA models before the observed epoch of the LEGA-C galaxy, with active star formation occurring in all objects at this redshift.  There is some tendency for the MaNGA galaxies to start their star formation later than the LEGA-C comparator.  This bias is partly a selection effect, in that any ``late blooming'' LEGA-C galaxies would not have been massive enough to observe in the LEGA-C survey, but it may also reflect the differing temporal resolution in the two data sets, with the more recent star formation in LEGA-C better able to resolve the earlier star-formation history than the MaNGA data in which all the early-forming stars are barely-distinguishable old populations at the current epoch.  Nonetheless, a consistent picture emerges of a later epoch of formation in these less-massive galaxies, reflecting the well-known downsizing effect \citep[e.g.][]{Panter2003,Kauffmann2003}. 
 It is interesting that this picture is also apparent in the images of the galaxies in  \autoref{fig:example_lowmass}: the LEGA-C galaxy shows a distinct spiral morphology consistent with a galaxy that is strongly star-forming, and while the present-day MaNGA galaxies still show some signs of star formation and a disk-like morphology, their appearance is significantly smoother, as might be expected as they evolve toward quiescence.

Finally, we come to the interesting galaxies that lie below the MZR. In \autoref{fig:example_massivelow}, we show a typical example of this class. 
The objects have a metallicity of around $0.5\,$dex below the usual mass--metallicity sequence. To check whether such low metallicity might be spuriously induced by degeneracy between the metallicity and other galaxy parameters, we have explored the posterior distribution of properties including gas-phase metallicity, averaged age, and dust optical depths of the young and old stellar population.  Among these distributions, the degeneracy between the average age of the stellar population and the current metallicity is found to be strongest. However, the amplitude of even this well-known age-metallicity degeneracy is found to be modest (typically $\sim0.1\,$dex), and could not be responsible for a variation of $0.5\,$dex in metallicity. Interestingly, no significant degeneracy is found between dust and stellar population properties, presumably because the SED from UV to NIR gives enough information to determine the attenuation reliably. We thus conclude that the measured low metallicities cannot simply be an artefact of degeneracies in the fitting process. However, the model fit implies a strong outflow with wind parameter $\lambda=5.5$ in the galaxy. Such a strong wind hints at a different underlying physical process, which will be explored in more detail in \autoref{sec:origins}. 

Focusing on the image of the LEGA-C galaxy, we see that it is rather smooth, consistent with the derived star-formation history, which has largely decayed away at the point of observation.  The matched MaNGA galaxies generally show a similar pattern of a decayed initial burst of star formation at this epoch, although again with the issue that the LEGA-C has better temporal resolution to resolve the sharpness of any initial burst.  In all cases, the metallicity remains at a depressed level, whose cause we can track down in the model as arising from a strong wind expelling enriched matter from these objects. 
What is notable about the MaNGA galaxies is that, following the LEGA-C epoch, all undergo a secondary phase of star formation accompanied by a rapid rise in metallicity, so that they end up on the MZR by the present day. Again, we can identify the physical cause of this effect in the model: it occurs when the winds in these systems cut off, allowing them to retain more gas to produce stars and more of the metals they generate to pollute their environment. This late star formation is also apparent in the images of the MaNGA galaxies, which show significant spiral structure that is not apparent in their LEGA-C progenitor.

From these examples, we can see that the LEGA-C and MaNGA galaxies matched in the mass--metallicity plane at $z \sim 0.7$ are inferred to follow very similar evolutionary paths up to that point. We can therefore, with some confidence, equate the LEGA-C galaxies with the progenitors of the matched MaNGA galaxies, and use the subsequent evolution of the MaNGA galaxies to infer how the LEGA-C galaxies will evolve to the present day.

\subsection{Morphological evolution}
\label{sec:evolution}

\begin{figure}
    \centering
    \includegraphics[width=0.5\textwidth]{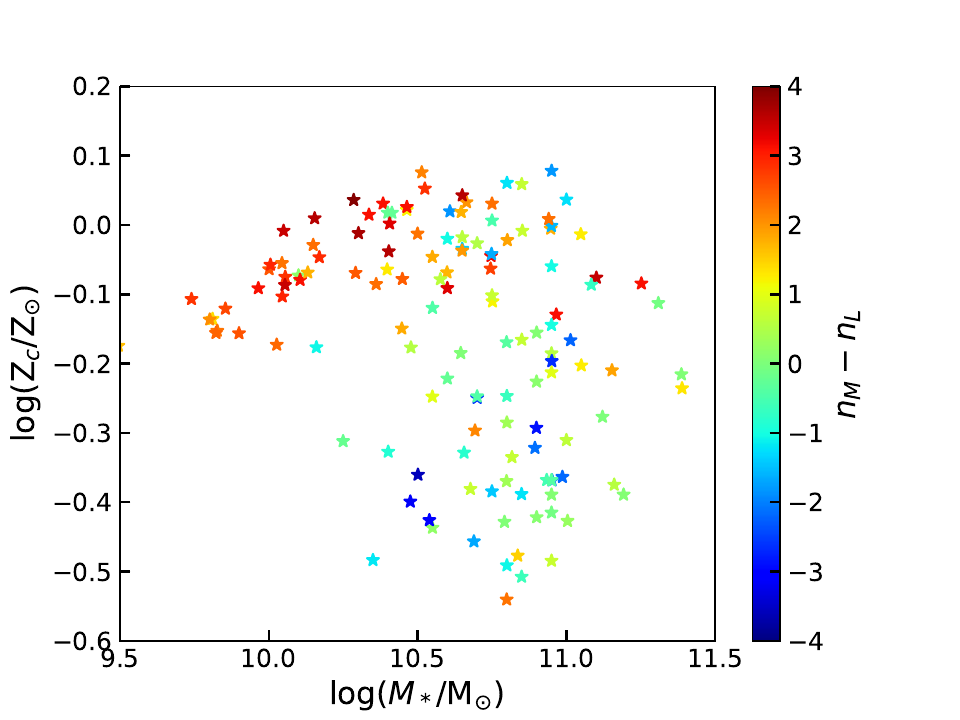}\\
     \caption{Differences between the observed Sersic indices in matched ``twins'' of LEGA-C and MaNGA galaxies on the mass--metallicity plane. Galaxies that become more bulge-like with time are red; those that become more disk-like are blue.
     }
     \label{fig:diff_sersic}
\end{figure}

\begin{figure}
    \centering
    \includegraphics[width=0.5\textwidth]{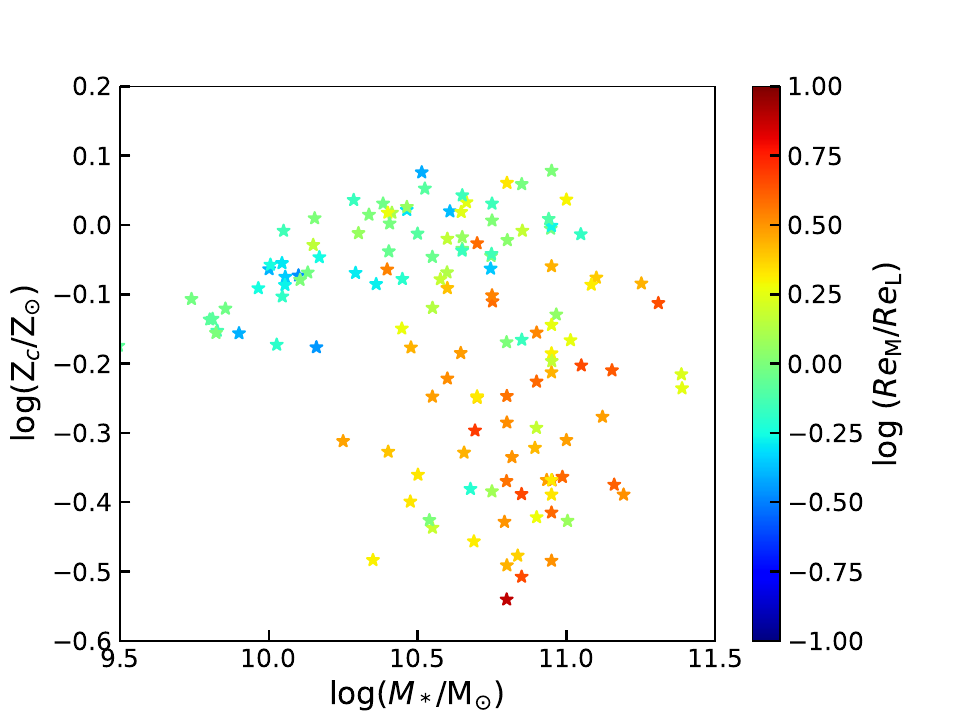}\\
     \caption{Ratios between the observed effective radii in matched ``twins'' of LEGA-C and MaNGA galaxies on the mass--metallicity plane. Galaxies that grow with time are red; those that shrink are blue.
     }
     \label{fig:diff_re}
\end{figure}

With the matched sets of galaxies, we can now look for other differences between their properties at $z \sim 0.7$ and the present day.  In particular, we can quantify the changes in morphology that seem to be apparent in the images that have been presented.  The simplest most robust comparison we can make involves the parameters of a single-component Sersic fit made to the galaxies' images, which return the Sersic index $n$ as a measure of central concentration and the effective radius $R_{\rm e}$ as a measure of their size. Such fits have already been made to the data presented here: for the MaNGA galaxies, we use values (denoted with a subscript ${\rm M}$ in what follows) from the NASA Sloan Atlas catalogue\footnote{\label{foot:nsa}\url{ http://www.nsatlas.org/}} (NSA, \citealt{Blanton2005}).  For the LEGA-C data, we use values (subscripted with an ${\rm L}$) provided by the data release catalogue, which are derived from the single component Sersic fits to the COSMOS HST/ACS images \citep{Scoville2007}.  For the MaNGA galaxies, we take the average of the five parameter values of the objects that have been matched to the LEGA-C galaxy, as described above.  In \autoref{fig:diff_sersic} we show the difference in Sersic index between the LEGA-C galaxies and their descendant MaNGA twins, and in \autoref{fig:diff_re}, we show the equivalent for the effective radius.

It is apparent from these figures that there has been significant evolution in galaxy morphology between these two epochs, but interestingly it is consistent with what we infer from the entirely independent analysis of different galaxies' evolutionary histories using the semi-analytic spectral fitting. As we have seen, massive high-metallicity galaxies completed their star formation significantly before even the LEGA-C galaxies were observed, and this quiescence is reflected in their morphologies, with neither Sersic index nor effective radius changing much between the LEGA-C galaxies and their MaNGA descendants.  At the low-mass end of the MZR, we have seen that star formation was ongoing to around the time of the LEGA-C observations, and only subsequently died away.  We can see this in the morphologies as the outer disk that had been actively forming stars and dominating the light fades away, so the Sersic index increased from the $n \sim 1$ that characterises a disk to the $n \sim 4$ of a more bulge-dominated system, and the whole system shrinks. In the high-mass low-metallicity systems, we see the opposite effect: the late star-formation in these systems occurs in their outer disks and comes to dominate their light, leading to a decrease in the Sersic index and an increase in their effective radius.

\subsection{The origin of the massive metal-poor galaxies}
\label{sec:origins}

The remaining substantive question is why some massive galaxies find themselves in the state where their metallicity remains low until after the epoch of the LEGA-C observations.  As we have seen from the models we have fitted to their spectra, the physical driver of this deficiency is that they maintain strong wind-driven mass-loss, which expels enriched material and so holds their metallicity at a low level.  Indeed, this mechanism has already been invoked to explain such properties -- \cite{Beverage2021} suggested that the low-metallicity galaxies they found in the LEGA-C data using more conventional spectral fitting could be explained by the effective removal of gas from these systems. The main question this raises is what the process is that maintains such strong wind effects in some galaxies but not others, and how all galaxies seem to know that they have to shut down these winds before the present day, to ensure that all systems now lie on the MZR.

One plausible candidate is environment -- one could envisage a scenario in which a galaxy which is a satellite to another system would be more likely to lose expelled material to its surroundings and companion than an isolated system of the same mass in which wind-driven material would more readily rain back down on the system that had tried to expel it. To test this possibility, we have made use of the wide variety of environment measures determined for COSMOS galaxies (of which the LEGA-C systems are a subset) by \cite{Darvish2017}. However, none of these environmental properties is found to significantly correlate with the location of the massive LEGA-C galaxies in the mass--metallicity plane. In addition, since environments are not likely to change substantially between the LEGA-C epoch and the present day, it is hard to understand how such an external agent can have ensured that the low-metallicity galaxies all undergo a rejuvenation in their star formation while none of the high-metallicity galaxies does.

We therefore look for an internal driver for this effect, where the natural candidate is AGN activity.  It is well established that strong AGN activity can drive gas out of a massive galaxy, which can have a significant impact on its star formation (see \citealt{King2015} for a review). For example, the IllustrisTNG simulations predict that galaxies more massive than $10^{10.6}{\rm M}_{\odot}$ have low metallicities at $z\sim1$ due to their strong AGN activity \citep{Torrey2019}. In our current scenario it is therefore possible that the wind term in the model is powered by an AGN, and the cut-off time parameter represents the point where AGN activity dies down.  In this picture, the low-metallicity galaxies are simply those whose AGN ``clock'' is running a little slow, and such activity has not shut down by the time the LEGA-C observations are made, but it subsequently does, allowing these galaxies to join the vast majority of non-active galaxies in the nearby Universe.  One implication of this scenario would be that the low-metallicity galaxies in LEGA-C should show systematically more signs of AGN activity than those already on the MZR.  Unfortunately, the limited spectral range of LEGA-C observations means that we cannot use the BPT diagram as a diagnostic of AGN activity.  

As an alternative indicator of AGN activity, one obvious choice is
the Mass Excitation (MEx) diagnostic proposed by \cite{Juneau2011}, which combines the stellar mass and the \hbox{[O\,{\sc iii}]}$\lambda$5007/H$\beta$ line ratio to 
distinguish between star formation and AGN activity -- these lines are both conveniently available in the LEGA-C data.  Unfortunately, however, this diagnostic was calibrated using the MZR from $z\sim0.1$ galaxies in SDSS \citep{Juneau2011}. For the current sample,
if a galaxy has a gas-phase metallicity well below the present-day MZR, the lower metallicity will induce higher \hbox{[O\,{\sc iii}]}$\lambda$5007/H$\beta$ \citep[e.g.][]{Kewley2008}, which could lead to it being misidentified as AGN using this diagnostic. Indeed, this bias may explain why \cite{Lewis2023} seems to find fewer massive metal-poor galaxies compared to this work -- they use the MEx diagram to exclude all galaxies that may have AGN contamination, and thus metal-poor galaxies are also more likely to be removed from their sample. 

\begin{figure}
    \centering
    \includegraphics[width=0.5\textwidth]{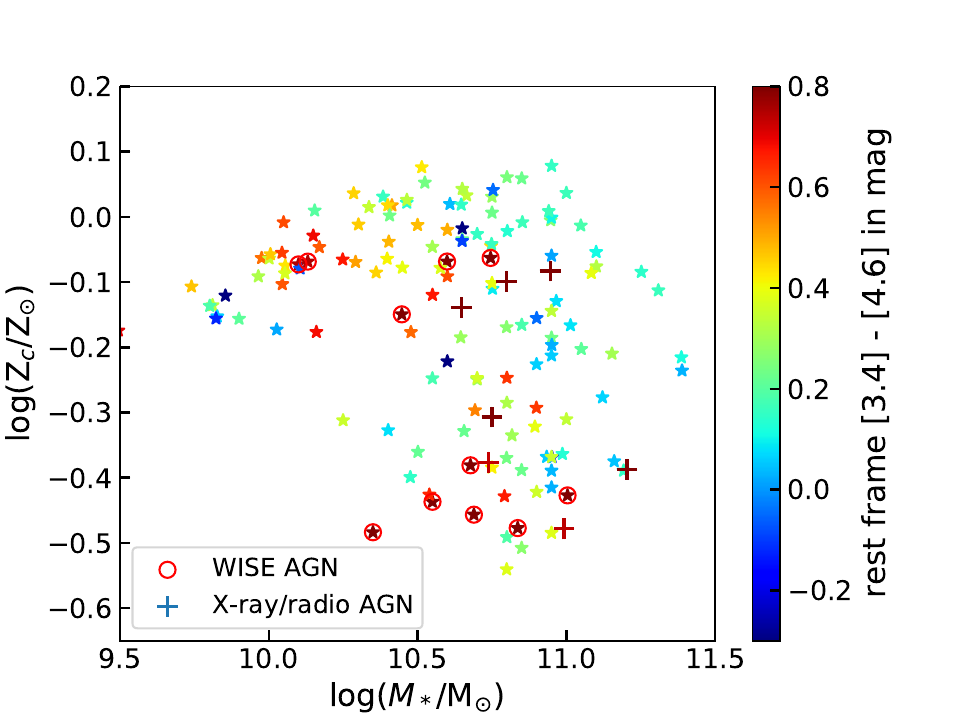}\\
     \caption{Rest frame $[3.4] - [4.6]$ colour of LEGA-C  galaxies on the mass--metallicity plane.  Those whose colours indicate AGN activity are highlighted with circles. Crosses show the locations of the seven objects identified as AGN from their X-ray or radio properties.
     }
     \label{fig:color_wise}
\end{figure}

Avoiding such emission-line-based methods, another alternative indicator is the infrared colour. Using the broad spectral range of the SED available for the LEGA-C data, we can infer the rest-frame colour between two WISE bands observed at 3.4 $\mu {\rm m}$ and 4.6 $\mu {\rm m}$, $[3.4] - [4.6]$. A plot of the mass--metallicity plane colour coded by this quantity is shown in \autoref{fig:color_wise}.  \cite{Wright2010} have shown that normal spiral and elliptical galaxies have $[3.4] - [4.6] < 0.7$, with contamination from AGN light shifting it to higher values.  Indeed, most of the massive galaxies in \autoref{fig:color_wise} have values below this cut-off. However, galaxies with higher values of $[3.4] - [4.6]$ are over-represented among the low-metallicity massive galaxies in this plot.

As a further check, we have also gone back to analyze the seven LEGA-C objects that were initially rejected from the sample because they had been flagged by their X-ray or radio properties as containing significant AGN activity.  Even after masking out the potentially AGN-contaminated parts of the SED, we have found that we are able to model the stellar components of these galaxies reasonably well.  As \autoref{fig:color_wise} shows, all of these objects lie below the MZR, with four at very low metallicity.  Taken together with the WISE colours, these results provide strong indications that persistent AGN activity provides an effective mechanism for delaying the enhancement of metallicity in massive galaxies, and it is only subsequent to the epoch of the LEGA-C observations that this activity dies down in all galaxies, allowing them to evolve to their present location on the mass--metallicity relation.

\section{Summary}
\label{sec:summary}
In this work, we directly model the formation and evolution of 152 galaxies at intermediate redshifts observed as part of the LEGA-C survey, using both the spectral and photometric data for these objects, and employ these models to place the galaxies reliably in the mass--metallicity plane. By carrying out a similar analysis of a sample of nearby galaxies from the MaNGA survey, we can reconstruct these objects' life histories, and hence determine where they should lie in the mass--metallicity plane in the epoch at which the LEGA-C galaxies were observed, thus matching the LEGA-C galaxies to their likely MaNGA descendants. The main results of this work are as follows:
\begin{itemize}

\item We find that many of the galaxies from the LEGA-C survey at $z\sim0.7$ already lie on the local mass--metallicity relation, indicating that the chemical enrichment process for these galaxies is largely complete. However, a significant fraction of relatively massive galaxies is found to lie at systematically lower metallicities by up to a factor of three.

\item
The archaeological reconstruction of MaNGA galaxies at $z\sim0.7$, when put through the same selection criteria as the LEGA-C galaxies, places them in the same locations in the mass--metallicity plane as the LEGA-C galaxies, providing some confidence in the reconstruction of their histories, and allowing us to pair LEGA-C galaxies individually with their likely descendants.

\item The matched galaxies show similar evolution up to the epoch of the LEGA-C observations, providing further confidence in both the modelling and the pairing of the galaxies.  The subsequent evolution of the MaNGA galaxies tells us what is likely to happen in the future of the LEGA-C galaxies. 

\item In massive high-metallicity galaxies, there is little subsequent evolution.  The main star formation occurred sufficiently long ago that there is not much change in morphology between the LEGA-C era and the present day either.

\item In lower-mass high-metallicity galaxies, the main burst of star formation occurred more recently, as expected from downsizing.  Star formation dies away after the LEGA-C epoch, resulting in morphological transformation as the star-forming disk fades leaving a more bulge-dominated appearance.

\item The anomalous high-mass low-metallicity galaxies are relatively quiescent at the epoch of the LEGA-C observations, but subsequently undergo a resurgence of star formation and boost in metallicity toward the mass--metallicity relation, which the modelling shows to be the result of galactic wind processes dying down.  This resurgence also shows up in the morphology, with the nearby examples of these galaxies showing the extended disk morphology associated with these new stars.

\item In seeking the origins of these high-mass low-metallicity galaxies, we note that a disproportionate number show indications of AGN activity.  It therefore seems plausible that these systems are ones in which such activity has persisted to somewhat lower redshift than average, with the associated winds suppressing the enhancement of metallicity.  By the present day, the remaining activity in these late developers will have regressed to the mean, allowing them to catch up with their delayed metallicity evolution, to match what we find in the MaNGA data.

\end{itemize}

\section*{Acknowledgements}
SZ, AAS and MRM acknowledge financial support from the UK Science and Technology Facilities Council (STFC; grant ref: ST/T000171/1). VMS acknowledges the FAPESP scholarships through the grants
2020/16243-3 and 2021/13683-5. 

We thank Jim Dunlop and Mike Edmunds for very insightful comments during a memorable meeting. 

For the purpose of open access, the authors have applied a creative commons attribution (CC BY) to any journal-accepted manuscript.

Funding for the Sloan Digital Sky Survey IV has been provided by the Alfred P. 
Sloan Foundation, the U.S. Department of Energy Office of Science, and the Participating Institutions. 
SDSS-IV acknowledges support and resources from the Center for High-Performance Computing at 
the University of Utah. The SDSS web site is www.sdss.org.

SDSS-IV is managed by the Astrophysical Research Consortium for the Participating Institutions of the SDSS Collaboration including the Brazilian Participation Group, the Carnegie Institution for Science, Carnegie Mellon University, the Chilean Participation Group, the French Participation Group, Harvard-Smithsonian Center for Astrophysics, Instituto de Astrof\'isica de Canarias, The Johns Hopkins University, Kavli Institute for the Physics and Mathematics of the Universe (IPMU) / University of Tokyo, Lawrence Berkeley National Laboratory, Leibniz Institut f\"ur Astrophysik Potsdam (AIP), Max-Planck-Institut f\"ur Astronomie (MPIA Heidelberg), Max-Planck-Institut f\"ur Astrophysik (MPA Garching), Max-Planck-Institut f\"ur Extraterrestrische Physik (MPE), National Astronomical Observatories of China, New Mexico State University, New York University, University of Notre Dame, Observat\'ario Nacional / MCTI, The Ohio State University, Pennsylvania State University, Shanghai Astronomical Observatory, United Kingdom Participation Group, Universidad Nacional Aut\'onoma de M\'exico, University of Arizona, University of Colorado Boulder, University of Oxford, University of Portsmouth, University of Utah, University of Virginia, University of Washington, University of Wisconsin, Vanderbilt University, and Yale University.

\section*{Data availability}
The data underlying this article were accessed from: SDSS DR17 \url{https://www.sdss.org/dr17/manga/}. The LEGA-C data is available in the third public data release (\citealt{vanderWel2021}), which can be accessed through the ESO Science Archive (\url{http://archive.eso.org/cms/eso-archive-news/Third-and-final-release-of-the-Large-Early-Galaxy-Census-LEGA-C-Spectroscopic-Public-Survey-published.html}). The derived data generated in this research will be shared on request to the corresponding author.

\bibliographystyle{mnras}
\bibliography{szhou} 

\appendix
\section{Influence of AGN on the KK04 metallicity calibration}
\label{appedix_caliration}
\begin{figure}
    \centering
    \includegraphics[width=0.5\textwidth]{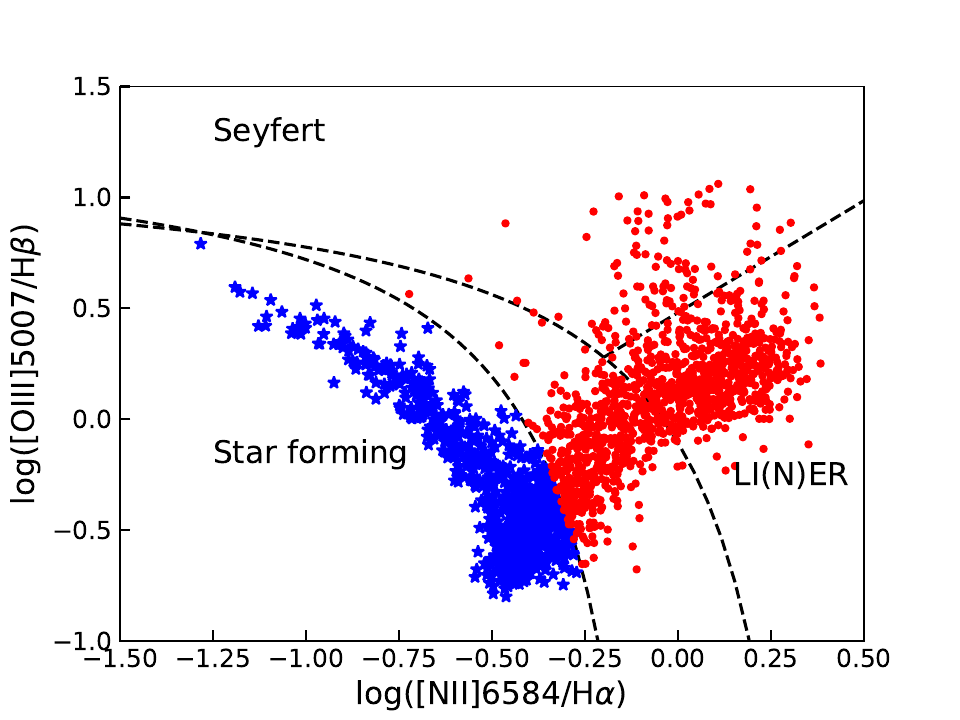}
     \caption{Distributions of the MaNGA sample galaxies on the BPT diagram. Dashed lines, from \citet{Kewley2001} and \citet{Kauffmann2003AGN}, divide the plane into regions of star-forming, Seyfert, and LI(N)ER galaxies.
     Pure star-forming galaxies are shown as blue stars, while composite/AGN galaxies are red dots.}
     \label{fig:BPT_MaGNA}
\end{figure}

We use our sample of MaNGA galaxies to assess the impact of possible contamination of AGN on the metallicity calculated via the KK04 approach. We obtain \hbox{[O\,{\sc iii}]}$\lambda$5007, \hbox{[N\,{\sc ii}]}$\lambda$6584, H${\alpha}$, and  H${\beta}$ emission line fluxes for the MaNGA sample from the DAP. In \autoref{fig:BPT_MaGNA}, we use the standard BPT diagram of emission line ratios log([O\,{\sc iii}]5007/H$\beta$) versus log([N\,{\sc ii}]6584/H$\alpha$) that is often used to identify AGN activity to divide the sample into pure star-forming galaxies (blue stars) and composite/AGN galaxies (red dots).

\begin{figure}
    \centering
    \includegraphics[width=0.5\textwidth]{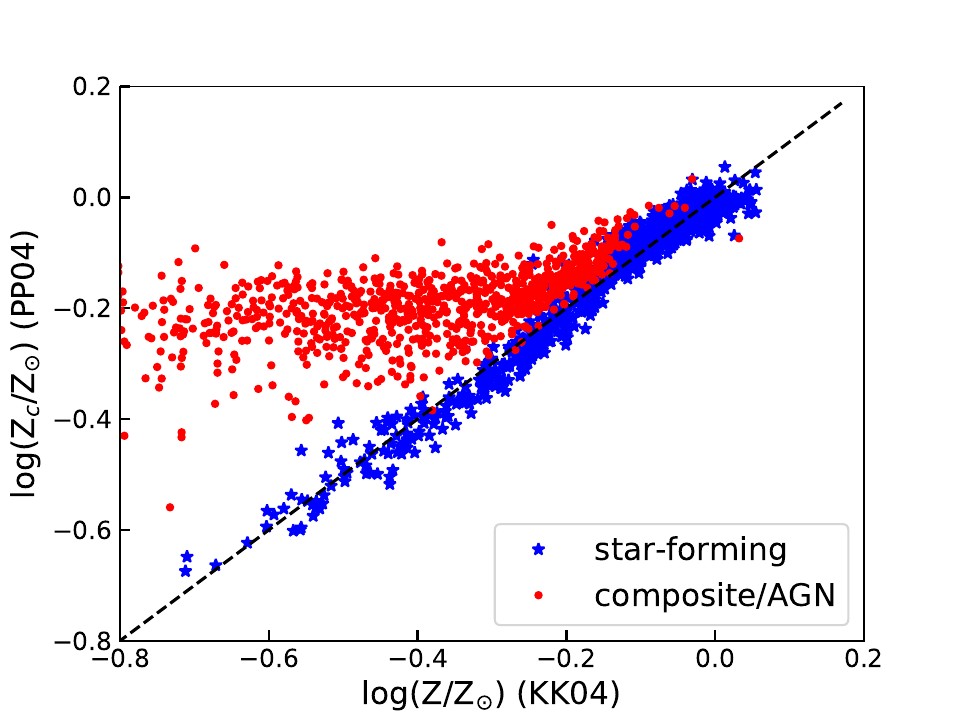}
     \caption{Comparison between metallicities derived from the 
     PP04 O3N2 approach and the KK04 calibration.  Pure star-forming galaxies are shown as blue stars, while composite/AGN galaxies are red dots. The dashed line indicates equality between the two measures.}
     \label{fig:compare_metal}
\end{figure}

 For each MaNGA galaxy, we then use both the PP04 O3N2 and the KK04 calibration to estimate its gas-phase metallicity. 
 In the calculation, we adopt the upper branch of the KK04 calibration, as has been done elsewhere in this work. to allow for differences in calibration, we use the formula provided by \cite{Kewley2008} to convert the KK04 metallicity into values that can be directly compared with the PP04 O3N2 metallicity. Both metallicities are scaled using solar abundance ratios, and are compared with each other in \autoref{fig:compare_metal}. From this plot, we see that for pure star-forming galaxies the two calibrations are in good agreement between $-0.8<\log(Z_{\rm g}/Z_{\odot})<0.2$. However, for composite/AGN galaxies, the two calibrations only match at the high metallicity end. When $\log(Z_{\rm g}/Z_{\odot})\lesssim-0.2$, they start to diverge from each other, such that the metallicities inferred from the KK04 calibration are biased towards lower metallicities. This is not unexpected, as the AGN produce extra [O\,{\sc iii}] emission, inducing a higher [O\,{\sc iii}]/H$\beta$ that corresponds to a lower metallicity in the upper branch of the KK04 calibration. As the AGN only have a minor impact on the [N\,{\sc ii}] emission, the PP04 O3N2 approach has been shown to be more reliable in LI(N)ER galaxies \citep{Kumari2019}. For the LEGA-C galaxies studied here, we do not have access to the measures required for the PP04 method, so we are stuck with the biased KK04 approach. Fortunately, as shown in \autoref{fig:compare_metal}, the impact of AGN contamination is only significant when the derived metallicity is lower than $\log(Z_{\rm g}/Z_{\odot})\lesssim-0.2$, and even then it always biases the result toward lower metallicity values. Accordingly, we  can assume that if the derived metallicity $\log(Z_{\rm g}/Z_{\odot})>-0.2$ then the constraints from PP04 metallicities are reliable.
 Otherwise, we simply adopt the value as a lower limit to the possible gas-phase metallicity. As discussed in the paper, even this conservative upper limit provides a useful unbiased constraint when fitting the full chemical evolution model to the data.

\section{Robustness of the fitting process}
\label{appedix_mocktest}
In this appendix, we aim to address the extent to which we can recover the stellar population properties, particularly the current gas-phase metallicity, from the constraints available in the LEGA-C dataset. To do this, we perform a set of tests analogous to those carried out in Paper~I.

As a starting point, we
generate a simulated sample with the procedure below. We first randomly generate a set of model parameters from the prior distribution listed in \autoref{tab:paras}. Corresponding mock SFH and ChEH are then calculated  using the chemical evolution models.
We exclude unreasonable parameter sets that would induce a current  gas-phase metallicity larger than 0.05, as covered by the SSP models. SFH and ChEH from valid parameters are sent to {\tt BIGS} to calculate a simulated spectrum using the BC03 SSP templates, redshifted to $z=0.7$. We then use filter response functions of the 28 photometry bands to calculate the SED of the simulated galaxy from this spectrum. Finally, we add Gaussian noise to the part of the spectrum in the LEGA-C spectral range such that its SNR is 20 {\AA}$^{-1}$, while the SNR of SED bands are uniformly set to 50. By repeating this process, we generate a sample containing 1000 simulated spectra and SEDs, which are then fitted using three fitting approaches:
\begin{itemize}
    \item Approach 1 fits a simulated galaxy with full information including the spectrum within the LEGA-C observation range (6300{\AA} to 8700{\AA}), the SED from 28 photometry bands, and the current gas-phase metallicity;
    \item Approach 2 fits a simulated galaxy with its spectrum and SED but does not include the current gas-phase metallicity;
    \item Approach 3 fits a galaxy with only its optical spectrum.        
\end{itemize}

\begin{figure*}
    \centering
    \includegraphics[width=1.0\textwidth]{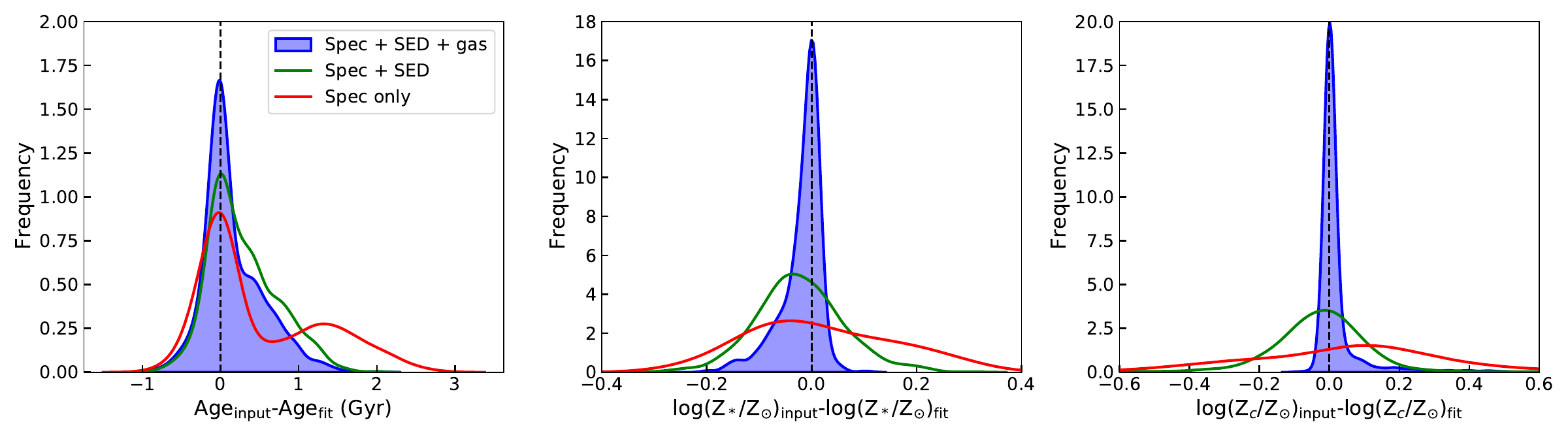}\\
     \caption{The distribution of the difference in the mean age (left), averaged stellar metallicity (middle) and current metallicity (right) between the best-fit model and input values. Results obtained from the three approaches using different constraining data are shown with different colours as indicated. Probability distributions are each normalized to 1.}
     \label{fig:mock_LEGAC}
\end{figure*}

After obtaining all the results, we determine the accuracy of the fitting by comparing the derived galaxy properties to the input models. Such comparisons are made for the specific properties of interest including the averaged stellar age and metallicity, as well as the current gas-phase metallicity. \autoref{fig:mock_LEGAC} shows how well the three approaches are able to reproduce these properties.  While the full fitting including gas-phase metallicity does a very good job of measuring all these properties, the limited range of the spectral data means that on its own it does a rather poor job, particularly on measures of metallicity. However, with additional information from the SED included, these uncertainties are significantly reduced, and much of the power given by the gas-phase metallicity is restored.

One further source of uncertainty lies in the model templates used in the fitting process.  As a test of their potential impact, we fitted the model data generated from the BC03 templates using the eMILES models \citep{Vazdekis2016}. Using the spectra and SED as constraints, we ascertained that the ages obtained using the eMILES models are as good as those from the BC03 models. Both averaged and current metallicities derived from the eMILES models have larger scatters, and a non-Gaussian distribution, indicating some systematic difference between the two models. However, the overall difference is only around 0.3$\,$dex, so does not significantly affect the results presented in this work.

As a final check of the reality of the low-metallicity systems that we have found in the LEGA-C data, we have checked if we are able to obtain an acceptable fit while imposing a high gas-phase metallicity as a constraint. We find that such a process cannot fit simultaneously both the spectral and the SED data.

\bsp	
\label{lastpage}
\end{document}